\documentclass[pra,preprint,groupedaddress,floatfix]{revtex4-1}
\usepackage[latin1]{inputenc}
\usepackage[greek, frenchb, english]{babel}
\usepackage{amsmath}
\usepackage{amsfonts}
\usepackage{graphicx}
\usepackage{epstopdf}
\usepackage{subfigure}
\usepackage{array}
\usepackage{color}
\usepackage{setspace}
\usepackage{amssymb,mathrsfs}
\usepackage{bm}

\graphicspath{{./figures/}}

\begin{document}


\title{Leidenfrost drops on a heated liquid pool}



\author{L. Maquet}
\email{lmaquet@doct.ulg.ac.be}
\affiliation{GRASP, CESAM, Universit\'e de Li\`ege, Li\`ege, Belgium}
\author{B. Sobac}
\email{bsobac@ulb.ac.be}
\affiliation{TIPs, Universit\'e libre de Bruxelles, C.P. 165/67, Brussels, Belgium}
\author{B. Darbois-Texier}
\affiliation{GRASP, CESAM, Universit\'e de Li\`ege, Li\`ege, Belgium}
\author{A. Duchesne}
\affiliation{GRASP, CESAM, Universit\'e de Li\`ege, Li\`ege, Belgium}
\author{M. Brandenbourger}
\affiliation{GRASP, CESAM, Universit\'e de Li\`ege, Li\`ege, Belgium}
\author{A. Rednikov}
\affiliation{TIPs, Universit\'e libre de Bruxelles, C.P. 165/67, Brussels, Belgium}
\author{P. Colinet}
\affiliation{TIPs, Universit\'e libre de Bruxelles, C.P. 165/67, Brussels, Belgium}
\author{S. Dorbolo}
\affiliation{GRASP, CESAM, Universit\'e de Li\`ege, Li\`ege, Belgium}


\date{\today}

\begin{abstract}
We show that a volatile liquid drop placed at the surface of a non-volatile liquid pool warmer than the boiling point of the drop can experience a Leidenfrost effect even for vanishingly small superheats. Such an observation points to the importance of the substrate roughness, negligible in the case considered here, in determining the threshold Leidenfrost temperature. A theoretical model based on the one proposed by Sobac~\textit{et~al.}~[Phys. Rev. E \textbf{90}, 053011 (2014)] is developed in order to rationalize the experimental data. The shapes of the drop and of the substrate are analyzed. The model notably provides scalings for the vapor film thickness. For small drops, these scalings appear to be identical to the case of a Leidenfrost drop on a solid substrate. For large drops, in contrast, they are different and no evidence of chimney formation has been observed either experimentally or theoretically in the range of drop sizes considered in this study. Concerning the evaporation dynamics, the radius is shown to decrease linearly with time whatever the drop size, which differs from the case of a Leidenfrost drop on a solid substrate. For high superheats, the characteristic lifetime of the drops versus the superheat follows a scaling law that is derived from the model but, at low superheats, it deviates from this scaling by rather saturating.
\end{abstract}

\pacs{47.55.D-,47.55.dp,47.55.pb}

\maketitle

\newcommand{\thickness}{(h_f)}
\newcommand{\dropPos}{h}
\newcommand{\poolPos}{e}

\section{Introduction}

The intense evaporation of a drop close to a hot surface can generate an insulating vapor layer which prevents boiling and over which the drop levitates, thereby featuring a high mobility~\cite{leidenfrost1756aquae,biance2003leidenfrost}. This so-called Leidenfrost effect was studied extensively in the middle of $19^{\rm th}$ century by the pharmacist Boutigny who considered Leidenfrost drops as a new state of matter called the spheroidal state and deemed it essential to describe the birth of the universe~\cite{boutigny1847nouvelle}. Far from being crucial in this regard, there is a renewal of interest nowadays for the Leidenfrost effect because of self-propelling behaviors~\cite{dupeux2011viscous}, nanofabrication possibilities~\cite{abdelaziz2013green} and chemical reactor capabilities~\cite{schwenzer2014leidenfrost}. Moreover, the Leidenfrost effect still leaves several important questions open such as that of the friction force experienced by these drops~\cite{quere2013leidenfrost}.

The levitation occurs when the temperature of the substrate exceeds a given temperature. The prediction of this critical Leidenfrost temperature is still an open question too, and even its definition is still subject to debate. Indeed, it is sometimes defined as the minimal temperature of the substrate for the Leidenfrost effect to occur~\cite{bernardin1999leidenfrost}, or as the temperature for which the lifetime of the drops is the largest~\cite{biance2003leidenfrost}. Recently, it has been proved that the texturation of solid surfaces strongly impacts the value of the Leidenfrost temperature~\cite{2011_kim}. Vakarelski~\textit{et~al.} showed that when a hot textured and superhydrophobic sphere is immersed in a water pool at $22^\circ$C, a vapour film is maintained until the sphere reaches the pool temperature without transition to nucleate boiling~\cite{vakarelski2012stabilization}. Kwon~\textit{et~al.} reported that sparse surface textures increase the Leidenfrost point because they promote droplet wetting via capillary wicking~\cite{kwon2013increasing}. Finally, Kruse and coworkers found that the Leidenfrost point can be increased by 175$^\circ$C thanks to a femtosecond laser surface processing~\cite{kruse2013extraordinary}. Thus, should one want to favor the Leidenfrost effect, the ideal substrate that one could imagine is molecularly smooth. Accordingly, a liquid substrate appears to be a good candidate. Few studies already considered the problem of a drop levitating in a Leidenfrost state above a hot liquid pool. However, these studies were restricted to liquid nitrogen drops~\cite{kim2006floating,snezhko2008pulsating} and none of them were focused on the question of the critical Leidenfrost temperature under these conditions. 

\begin{figure}[h!]
\centering
	\includegraphics[width=8cm]{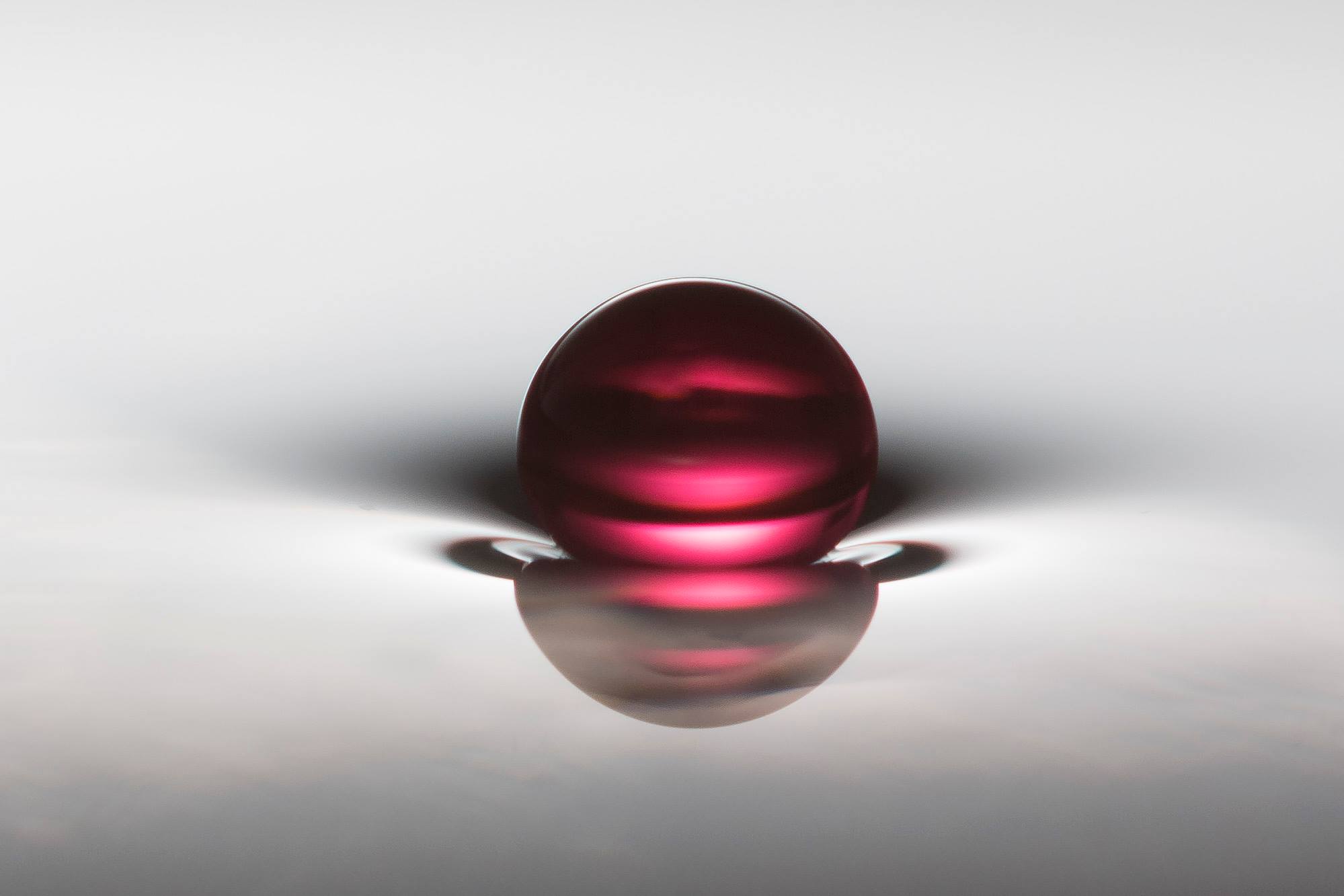}
	\caption{An ethanol drop with red dye levitates in the Leidenfrost state over a pool of silicone oil V20. The drop radius is $R=1.2$ mm and the pool is at $T_p =80^\circ$C while the boiling point of ethanol is $T_{\rm sat} = 78^\circ$C. (Image credit: Florence Cavagnon)}
	\label{figures/photoFlo}
\end{figure}

In this paper, we are concerned with a Leidenfrost drop placed above a hot liquid pool. Our goal is to broadly tackle this problem by addressing the question of the critical Leidenfrost temperature in this situation and providing a better understanding as for the interaction between the drop and its liquid substrate. In parallel, the differences between liquid and solid substrates will be highlighted as far as the Leidenfrost effect is concerned. 

First, we describe the experimental setup and the developed model in section~\ref{sec:methods}. In section~\ref{sec:results}, the main results obtained are presented, the experimental findings being confronted with the model. Then, in section~\ref{sec:considerations}, some details about the limitations of our analysis as well as the associated perspectives are discussed. Finally, conclusions are summarized in section~\ref{sec:conclusion}.

\section{Methods: experimental and theoretical }\label{sec:methods}

\subsection{Experimental}\label{sec:setup}

\begin{figure}[h!]
\centering
		\includegraphics[width=8cm]{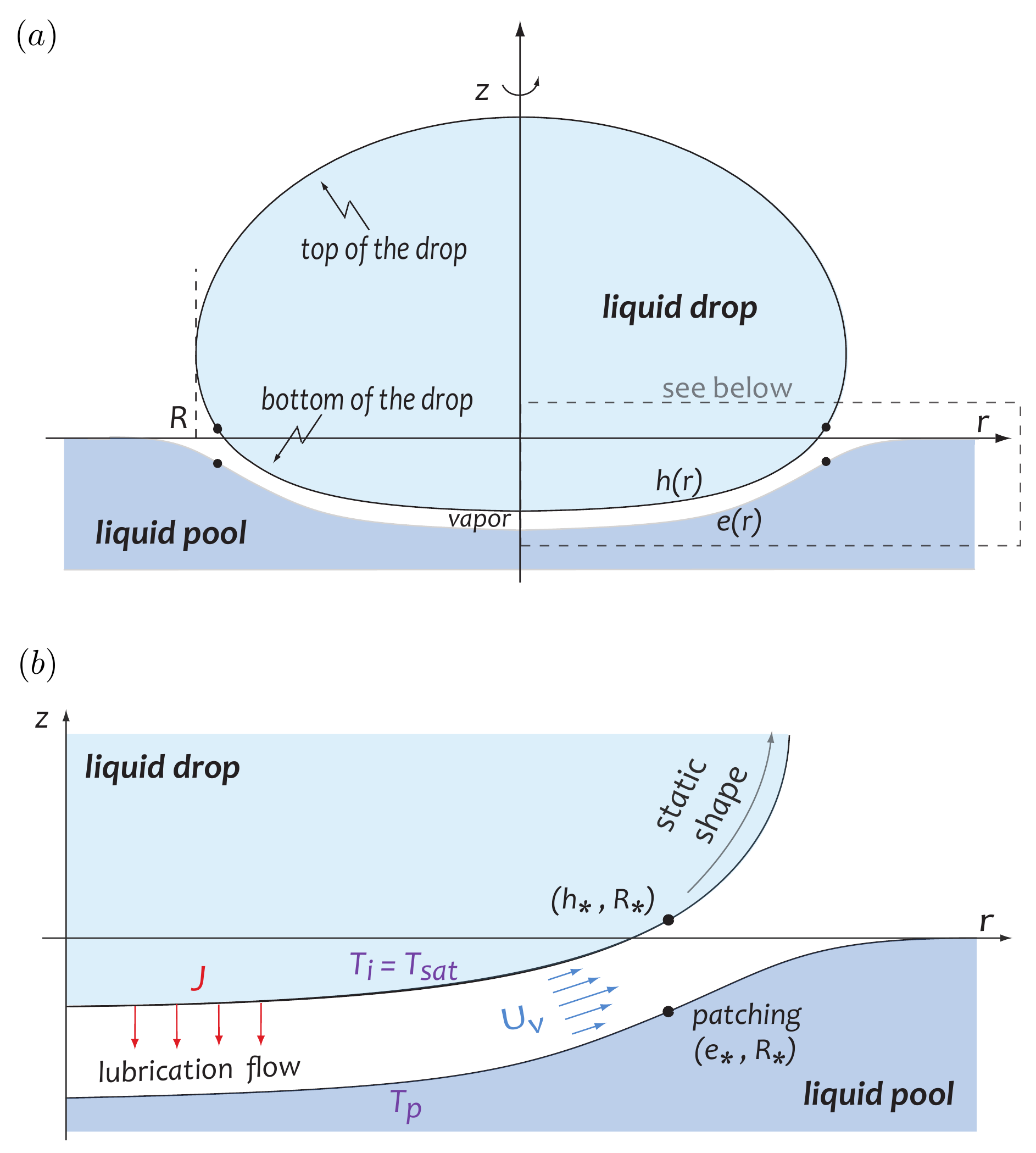}
 	\caption{Schematic illustration of a Leidenfrost drop levitating on a liquid pool. Figure (a) gives a large view while figure (b) shows a zoom on the region of the vapor film.}
		\label{fig:setup_theory}
\end{figure}

The experimental setup is the following: a drop of a radius $R$ (as viewed from above) is released above a hot pool of another liquid (see Fig.~\ref{fig:setup_theory} and supplementary material). The pool is maintained at a temperature $T_p$ thanks to a heating plate, a PID controller and a thermocouple that is immersed in the pool close to its surface. The temperature uncertainty is $\pm 2^\circ$C coming from the precision of the device and small variations that can occur during the evaporation of one drop. The boiling point of the drop is denoted as $T_{\rm{sat}}$. The difference between the temperature of the pool and the boiling temperature of the liquid of the drop is denoted as $\Delta T = T_p  - T_{\rm{sat}}$ and will always be positive throughout this paper. Thus, $\Delta T$ will be called the superheat. Experiments were carried out with drops released at ambient temperature and at $T_{\rm{sat}}$, showing no difference of behavior except a better stability at low superheats.

The density, surface tension and dynamic viscosity are denoted respectively by $\rho_d $, $\gamma_d $ and $\mu_d $ for the drop liquid and $\rho_p $, $\gamma_p $ and $\mu_p $ for the liquid of the pool. Thus, we introduce the capillary length of the drop as $\ell_{c_d}=\sqrt{\gamma_d/\rho_d g}$ and that of the pool as $\ell_{c_p}=\sqrt{\gamma_p/\rho_p g}$, where $g$ is the gravity acceleration.

The study focuses on ethanol drops over silicone oil pools. Indeed, silicone oil enables heating the pool up to roughly $200^\circ$C and is practically non-volatile as compared to other liquids at these temperatures. The boiling temperature of ethanol is $T_{\rm{sat}} = 78^\circ$C. That allows us to work with superheats ranging from $0^\circ$C to about $100^\circ$C. The depth, length and width of the liquid pool are more than ten times larger than the drop radius and can be thus considered as practically infinite. Indeed, no difference in behavior has been observed for different sizes of the container as soon as the depth remains larger than the deformation of the pool surface. 

Two types of measurements were taken. First, we captured the deformations of the surface of the pool from the side with a telecentric lens. These experiments validate the model developped hereafter. Second, we recorded the maximum radius of the drop from the top as a function of time in order to quantify the evaporation rate of the drops. 


\subsection{Theoretical}\label{sec:Model}

In order to model the experimental phenomena, we consider an axisymmetric drop levitating above a hot liquid pool (see Fig.~\ref{fig:setup_theory}). The liquid pool is assumed to be non-volatile and isothermally maintained at a temperature $T_p$ larger than the boiling point of the drop $T_{\rm{sat}}$. The size of the drop is measured by its maximum radius $R$, as seen from above. The vertical position $z$ of the drop bottom and pool interfaces, which are functions of the radial coordinate $r$, are denoted as $\dropPos(r)$ and $\poolPos(r)$, respectively. The difference of these two quantities, namely the thickness of the vapor film, is denoted $h_f(r) = h(r) - e(r)$. The level $z=0$ is chosen at the unperturbed liquid pool surface, expected to be attained far away from the drop, and hence $e \rightarrow 0$ as $r \rightarrow \infty$. Notationwise, note that here and throughout this paper, the presence or the absence of a tilde is meant to distinguish between dimensionless and dimensional quantities, respectively. Lengths are made dimensionless by the capillary length of the drop liquid $\ell_{c_d}$, whereas the excess pressure quantities to be introduced below are adimensionalized with the scale $\gamma_d / \ell_{c_d}=\rho_d g \ell_{c_d}$. As the evaporation of Leidenfrost drops is typically long compared to thermal and viscous relaxation times, we are looking for steady shapes only (quasi-steady approximation). 

Concerning the shape of the drop, the model proposed is based on the one presented in Sobac~\textit{et~al.}~\cite{sobac2014}. Two different regions are distinguished. The upper part of the drop, above the vapor film (beyond the patching point shown in Fig. \ref{fig:setup_theory}), is assumed to be an equilibrium shape, for which the Laplace pressure $\gamma_d \kappa$ (with $\kappa$ the curvature of the drop surface) locally balances (up to a constant) the hydrostatic pressure $-\rho_d g z$. In dimensionless form, this simply reads

\begin{equation}
\tilde \kappa+(\tilde z-\tilde z_{\rm top})=\tilde \kappa_{\rm top} \ ,
\label{TopEquilibriumDropShape_equation}
\end{equation}

\noindent where $\tilde \kappa_{\rm top}$ is the dimensionless curvature at the top of the drop ($\tilde z=\tilde z_{\rm top}$), which is here used as a free parameter controlling the drop size. For a given $\tilde \kappa_{\rm top}$, numerically integrating this differential equation ($\tilde \kappa$ being a function of the drop shape and its derivatives) from the symmetry axis yields the corresponding equilibrium shape and in particular the value of the maximum radius $\tilde R=\tilde R(\tilde \kappa_{\rm top})$. Note that the choice of $\tilde z_{\rm top}$ affects only the vertical position of the resulting equilibrium shape, but not the shape itself or the value of $\tilde R$ (which depends only upon $\tilde \kappa_{\rm top}$). This is important since the value of $\tilde z_{\rm top}$ remains unspecified at this stage and is eventually determined only together with the overall shape of the Leidenfrost drop.

This ``upper equilibrium drop" solution is assumed to be valid up to a point located at $\tilde r=\tilde R_\ast$, where non-equilibrium effects of evaporation and viscous pressure losses in the vapor flow start to progressively come into play. In this ``vapor layer region" $0<\tilde r<\tilde R_\ast$, the thickness of the vapor film and its slopes are small enough so as to use the lubrication approximation. Furthermore, the slopes of the drop bottom and pool interfaces are here both considered to be small. The gas itself is assumed to be composed of pure vapor (no air), incompressible and its properties are taken as constant. Then, ignoring possible motions inside the drop (following~\cite{2009_Snoeijer}), the excess pressure (over the ambient one) in the vapor film is found from the balance of forces normal to the drop surface as $P_v= \gamma_d \kappa_{\rm top}+\rho_{d} g (z_{\rm top} - \dropPos) -\gamma_d  \kappa$. This excess pressure  drives a Stokes flow with a volumetric flux $\vec{q}_v=-\frac{{\vec{\nabla} P_v}}{12\mu_v} (h-e)^3$, where $\mu_v$ is the vapor dynamic viscosity. Note the coefficient $1/12$ in the mobility factor, reflecting of no-slip conditions imposed at both the drop and liquid pool interfaces. Assuming that heat is only transferred by conduction across the film, the local evaporation flux at the interface is expressed as $\mathcal{J}={\mathcal{L}}^{-1}\lambda_v \Delta T  / (h-e)$, where $\lambda_v$ is the vapor thermal conductivity, $\mathcal{L}$ is the latent heat of vaporization, and $\Delta T=T_p-T_{\rm sat}$ is as before the superheat. Finally, the vapor mass conservation under the lubrication hypothesis reads $\vec{\nabla} \cdot \vec{q}_v - \mathcal{J}/\rho_v=0$ (assuming quasisteadiness), where $\rho_{v}$ is the vapor density. Combining these results, scaling all lengths with $\ell_{c_d}$ and assuming the axial symmetry yields the following equation :

\begin{equation}
\frac{1}{12} \frac{1}{\tilde r} \frac{\partial}{\partial \tilde r} \left( \tilde r (\tilde{h}-\tilde{e})^3 \frac{\partial}{\partial \tilde r}\left( \tilde \dropPos+ \tilde \kappa \right)\right)-\frac{\mathcal{\tilde E}}{(\tilde{h}-\tilde{e})}=0 \ ,
\label{lubrication_equation}
\end{equation}

\noindent with an evaporation number $\mathcal{\tilde E}$ defined by

\begin{equation}
\label{evap_number}
\mathcal{\tilde E}=\frac{\lambda_v \mu_v \Delta T}{\gamma_d \rho_v \ell_{c_d} \mathcal{L}} \ .
\end{equation}

\noindent As for the curvature $\tilde \kappa$, it is here left as a full-form expression

\begin{equation}
\tilde \kappa=\frac{\frac{\partial^2 \tilde \dropPos}{\partial \tilde r^2}+\frac{1}{\tilde r} \left(1+\left({\frac{\partial \tilde \dropPos}{\partial \tilde r}}\right)^2\right)\frac{\partial \tilde \dropPos}{\partial \tilde r}}{\left(1+{\left(\frac{\partial \tilde \dropPos}{\partial \tilde r}\right)}^2\right)^{3/2}} \ ,
\label{curvature}
\end{equation}

\noindent with no small-slope simplifications applied. The reason is that such a full expression proves to eventually provide for a smoother numerical matching with the upper part of the drop. Appart from the size $\tilde R$ of the drop, $\mathcal{\tilde E}$ is an other parameter of the dimensionless problem and depends on the fluid properties and on the superheat $\Delta T=T_p-T_{\rm sat}$. While the liquid properties are taken at $T_{\rm sat}$, the vapor properties are here  evaluated at the mean temperature of the vapor film, $(T_p+T_{\rm sat})/2$. Four boundary conditions are needed to supplement Eqs~(\ref{lubrication_equation}) and (\ref{curvature}): symmetry conditions at ${\tilde r}=0$, \textit{i.e.} ${\tilde \dropPos}'(0)=0$ and ${\tilde \kappa}'(0)=0$, while at ${\tilde r}=\tilde R_\ast$ the solution must match with the earlier obtained upper equilibrium shape of the drop, \textit{i.e.} we require the continuity of $\tilde \dropPos'(r)$ and of $\tilde \kappa(r)$. The continuity of $\tilde \dropPos(r)$ itself here merely amounts to finding the appropriate vertical shift of the upper equilibrium shape, \textit{i.e.} to determining the value of $\tilde z_{\rm top}$ (cf. above).

The liquid pool surface shape is assumed to be predominantly governed by a balance between hydrostatic and capillary pressures, as well as by the pressure exerted by the vapor film. In dimensionless form, this reads

\begin{equation}
\frac{\partial^2 \tilde \poolPos}{\partial \tilde r^2}+\frac{1}{\tilde r} \frac{\partial \tilde \poolPos}{\partial \tilde r}- \frac{\ell_{c_d}^2}{\ell_{c_p}^2} \tilde \poolPos=\frac{\gamma_d}{\gamma_p} \tilde P_v \ ,
\label{eq_pool}
\end{equation}

\noindent where the curvature expression is here written assuming small slopes. The dimensionless excess pressure in the vapor film is given by $\tilde P_v=\tilde \kappa_{\rm top}+\tilde z_{\rm top}- \tilde h -\tilde \kappa$ at $\tilde r < \tilde R_\ast$. On the other hand, we expect $\tilde P_v=0$ at $\tilde r \geqslant \tilde R_\ast$. Note that $\tilde P_v$ thereby defined is actually continuous at $\tilde r=\tilde R_\ast$. Indeed, $\tilde \dropPos$ and $\tilde \kappa$ are continuous at $\tilde r=\tilde R_\ast$ (see the boundary conditions above), and $\tilde \kappa_{\rm top}+\tilde z_{\rm top}- \tilde h -\tilde \kappa$ must vanish where the equilibrium shape is attached, quite in accordance with Eq.(\ref{TopEquilibriumDropShape_equation}). For $\tilde r>\tilde R_{\ast}$, Eq.(\ref{eq_pool}) can thus be simplified to $\left(\frac{\partial^2 \tilde \poolPos}{\partial \tilde r^2}+\frac{1}{\tilde r} \frac{\partial \tilde \poolPos}{\partial \tilde r}\right)- (\ell_{c_d}^2 / \ell_{c_p}^2)\tilde \poolPos=0$, with an analytical solution $\tilde \poolPos=\mathcal{\tilde C} K_0\left(\frac{\ell_{c_d}}{\ell_{c_p}}\tilde r\right)$ satisfying the condition at infinity. Here $K_0$ is the modified Bessel function of the second kind and $\mathcal{\tilde C}$ is an unknown coefficient to be determined. In the interval $0<\tilde r<\tilde R_{\ast}$, the following boundary conditions are used to solve Eq.(\ref{eq_pool}) : the symmetry condition at ${\tilde r}=0$, \textit{i.e.} ${\tilde \poolPos}'(0)=0$, while at ${\tilde r}={\tilde R}_{\ast}$ the solution must match with the above analytical solution, \textit{i.e.} we require the continuity of $\tilde \poolPos(r)$ and of $\tilde \poolPos'(r)$. Note that it is indeed three boundary conditions that are needed here for the second-order differential equation, for $\mathcal{\tilde C}$ is also an unknown. 

The problem given in the interval $0<\tilde r<\tilde R_\ast$ by Eqs (\ref{lubrication_equation}), (\ref{curvature}), (\ref{eq_pool}) with the formulated boundary conditions is discretized in a standard way by second-order finite differences on a uniform grid. The resulting nonlinear algebraic system of equations for the values $\tilde \dropPos$, $\tilde \kappa$, $\tilde \poolPos$ at the grid points as well as the values of $\tilde z_{\rm top}$ and $\mathcal{\tilde C}$ is solved by the Newton-Raphson method. It is checked \textit{a posteriori} that the choice of the patching point $\tilde R_\ast$ has no significant influence on the results.\\

\section{Results and discussion}\label{sec:results}

\subsection{First observations}\label{sec:quali_obs}

If we release a drop of ethanol (boiling point $T_{\rm{sat}} =78^\circ$C) on an oil pool at $T_p < T_{\rm sat}$, it contacts the surface of the pool and form a liquid lens. However, if the temperature of the pool is above $T_{\rm sat}$, a drop like the one shown in Fig.~\ref{figures/photoFlo} can be observed. This figure shows an ethanol drop of $1.2$ mm in radius levitating over a pool of silicone oil V20 at $T_p =79^\circ$C. Erythrosine has been added to the ethanol for a better visualisation. We clearly distinguish that the drop stands above the deformed surface of a liquid pool and keeps an almost spherical shape. For larger drops, the deformation of the pool surface is observed to be more significant. Drop radii up to $5$ mm could be experimentally reached. However, for such large drops, even a small perturbation induces their contact with the pool. On the other hand, the large drops become more stable when larger superheats are used in the experiments. Over time, the radius of the levitating drop decreases for about a minute. When the drop becomes very small ($\sim 0.3$ mm), the drop contacts the liquid pool forming a liquid lens which quickly evaporates. Thus, the end of life of the drop is different between the cases of a liquid substrate and of a solid substrate for which a take-off behavior is observed~\cite{celestini2012take}.

Observing the Leidenfrost effect for a superheat as low as $\Delta T = 1^\circ$C is a feat that is never seen on a solid substrate but turns out to be possible on a liquid pool. Thus, in this sense, a liquid seems to offer an ideal substrate to study the Leidenfrost effect.

As in the case of a solid substrate, the drops are highly mobile due to the presence of the vapor film, at least for small-sized drops when the pool deformation is not too strong. Many parameters seem to influence such mobility, including the temperature of the pool and the radius of the drops. We also observe that the drops are repelled by positive menisci that emerge from the pool, \textit{e.g.} those induced by the walls of the container. On the contrary, they are attracted by objects that dig the surface of the pool as cereals do at the surface of a bowl of milk~\cite{vella2005cheerios}. Thus, when two drops are on the same pool, they attract each other until coalescing. Sometimes, it happens that such coalescence is delayed for a few seconds. \\

\begin{figure}[h!]
\centering
\begin{minipage}{0.45\textwidth}
\centering
(a)\\
\vspace{0.2cm}
\includegraphics[width=8cm]{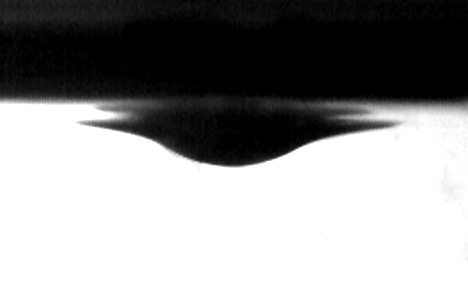}
\end{minipage}
\begin{minipage}{0.45\textwidth}
\centering
\vspace{-1cm}
(b)\\
\vspace{0.1cm}
\includegraphics[width=8cm]{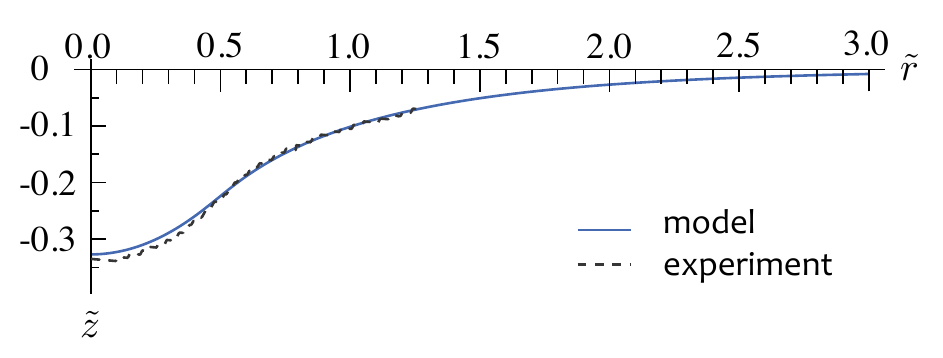}
\end{minipage}
 \caption{(a) Side view of the surface of the pool and (b) analysis of the deformations of the pool surface caused by the levitating drop. This experiment corresponds to an ethanol drop ($\ell_{c_d} =1.5 \, \rm{mm}$) on a silicone oil pool ($\ell_{c_p} =1.2 \, \rm{mm}$) at a temperature $T_p =118^\circ \textrm{C}$. The liquid density ratio is $\rho_d  /\rho_p =0.86$ and the drop radius $R = 1.2$ mm.}
\label{fig:pool_deformation}
\end{figure}

\subsection{Drop shapes and pool deformation}\label{sec:Mod_pool}

Figure \ref{fig:pool_deformation}(a) shows a direct experimental observation from the side of the deformation of the interface of the liquid pool below a drop of a radius $R = 1.2$ mm. From this type of images, the profile of the liquid pool interface can be extracted as illustrated in Fig. \ref{fig:pool_deformation}(b) for the same case (black dashed line). In this same figure, the profile of the liquid pool surface predicted by the model is also plotted for comparison. A rather good agreement is observed between the experiment and the fitting-parameter-free model.

\begin{figure}[h!]
\centering
		\includegraphics[width=9cm]{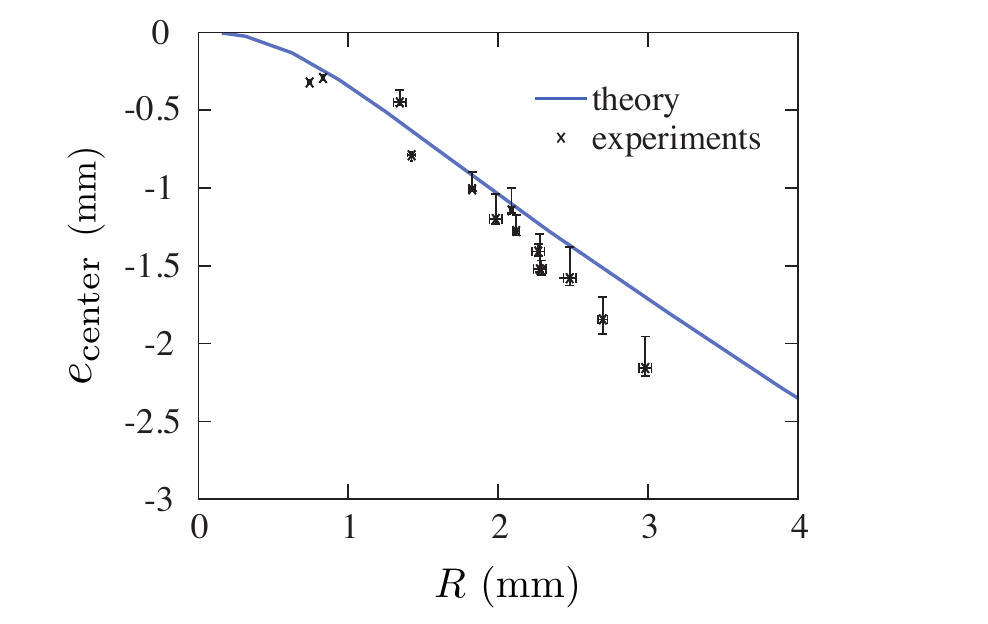}
 	\caption{Computed maximum depth of the silicone oil V20 pool deformation as a function of the ethanol drop size for $\mathcal{\tilde E}=2.23\times10^{-7}$ ($T_p=118^{\circ}$C for ethanol). The crosses represent experimental measurements.}
		\label{fig:poolDeformation}
\end{figure}

The maximum deformation $\poolPos_{\rm center}$ of the liquid pool at the symmetry axis is investigated as a function of the drop radius $R$. Figure~\ref{fig:poolDeformation} shows that the depth of the pool deformation monotonically increases with the drop size. Moreover, it highlights a reasonable agreement between the predictions of the model and the measurements. The measurements shown in Fig.~\ref{fig:pool_deformation} and Fig.~\ref{fig:poolDeformation} thus validate the model, which then allows obtaining data that are not accessible through our experiments such as the shape of the vapor layer (necessary to predict the evaporation rate of the drops).

\begin{figure*}[ht!]
\centering
		\includegraphics[width=\textwidth]{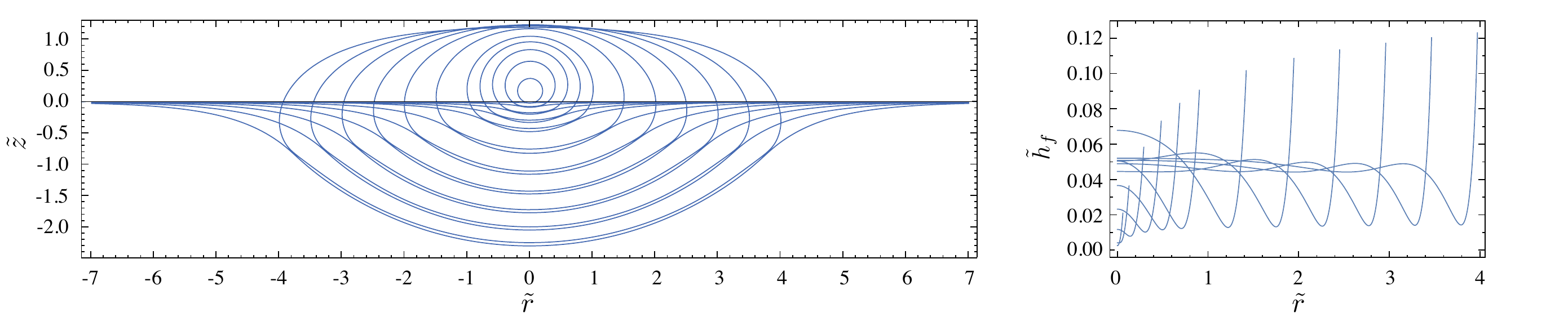}
 	\caption{Numerically determined shapes of (a) ethanol Leidenfrost drops on a liquid pool of silicone oil V20 and (b) their corresponding underlying vapor films for $\tilde R=0.2, 0.4, 0.6, 0.8, 1, 1.5, 2, 2.5, 3, 3.5, 4$, and $\mathcal{\tilde E}=2.23\times10^{-7}$ (which corresponds to $T_p=118^{\circ}$C for ethanol).}
		\label{fig:DropShapes}
\end{figure*}

Figure~\ref{fig:DropShapes}(a) shows typical shapes of ethanol Leidenfrost drops over a silicone oil pool computed by our model for various drop sizes. It is seen that smaller drops are quasi-spherical, whereas larger ones are flattened at the top by gravity. This transition between spheres and puddles occurs when gravity overcomes the capillary forces. Typically, the effect of gravity on the drop shape can be disregarded when the dimensionless drop radius $\tilde R$ is well smaller than one. At the same time, under the drop, the liquid pool adopts a concave shape which becomes increasingly marked with the drop size. The larger is the drop size, the deeper it is immersed in the pool (see also Fig.~\ref{fig:poolDeformation}).

Next, we numerically investigate the shape of the vapor film by considering the radial variation of its thickness $\tilde{h}_f$ in Fig.~\ref{fig:DropShapes}(b). $\tilde{h}_f$ is defined as the thickness of the vapor film projected perpendicularly to the pool surface as $\tilde{h}_f=(\tilde{h}-\tilde{e})/ \sqrt{1+{(\frac{\partial \tilde{e}}{\partial \tilde{r}})}^2}$. It is seen that the thickness profile forms a pronounced depression near the outer edge, referred to as the neck. The vapor layer can thus be view as composed of a vapor pocket at its center, surrounded by an annular neck region. The film thickness at the neck location appears to increase with the drop size. For $\tilde R \lesssim 1$,  the pictured profile of the vapor film thickness is similar to the one obtained in the situation of a Leidenfrost drop over a flat substrate~\cite{sobac2014}, which is not astonishing given the fact that the substrate is only weakly deformed in this case. For these small drop sizes, the depth of the vapor pocket increases with the drop size. However, for $\tilde R\gtrsim1$, the depth of the vapor pocket gets slightly lower and then reincreases. For these large drop sizes, the shape of the vapor pocket proves to become more complex, somewhat flattened and wavy . 

It is important to notice that calculations have been performed until $\tilde R=10$ and no sign of chimney formation has been observed until then. In contrast, for Leidenfrost drops over a flat substrate, the vapor pocket depth is found to increase sharply as $\tilde{R}\sim4$ is approached, and no meaningful solution is obtained for larger $\tilde{R}$~\cite{sobac2014}, which must actually correspond to a chimney formation by a Rayleigh-Taylor-like mechanism~\cite{biance2003leidenfrost,2009_Snoeijer}. Experimentally, the maximum drop radius that we could observe on a liquid pool is $\tilde R=3.5$, which was achieved by injecting liquid in a drop already stable on the pool. Above that radius, even though no sign of chimney is observed, the drops tend to contact the surface of the pool due to perturbations such as capillary waves at their surfaces.

\begin{figure}[h!]
\centering
		\includegraphics[width=9cm]{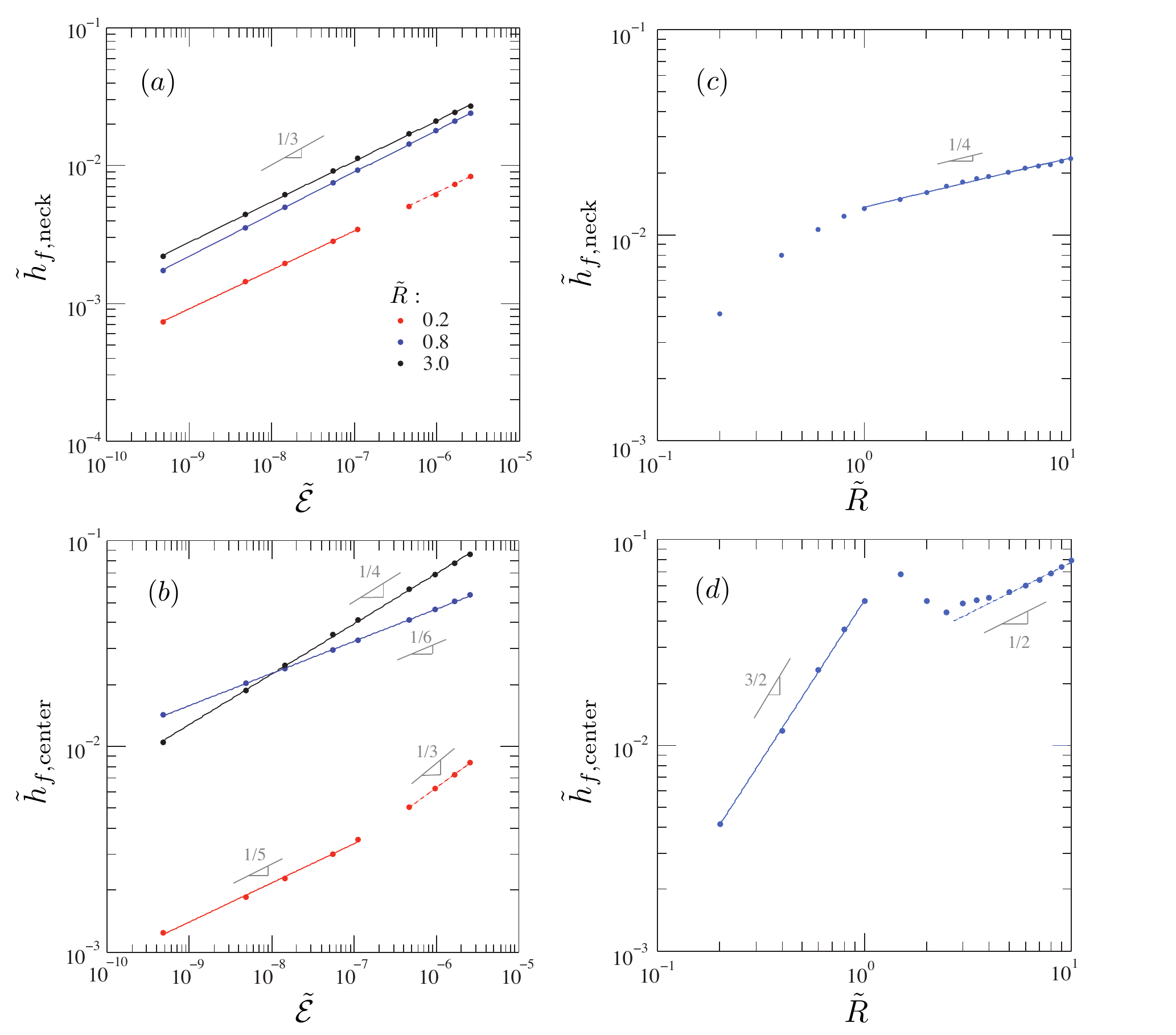}
 	\caption{Dimensionless vapor film thickness (a) at the neck and (b) at the center as a function of the evaporation number for various drop sizes. [(c) and (d)] The same versus the drop size for an evaporation number $\mathcal{\tilde E}=2.23\times10^{-7}$ ($T_p=118^{\circ}$C). Dimensional thickness values can be restored by multiplying by the ethanol capillary length $\ell_c=1.56$ mm.}
		\label{fig:scaling}
\end{figure}

Figure~\ref{fig:scaling} presents characteristic dimensions of the vapor film between the Leidenfrost drop and the pool surface. It shows the vapor film thicknesses at the neck $\tilde{h}_{f,\rm neck}$ and at the center $\tilde{h}_{f,\rm center}$ as a function of the evaporation number and the drop size. One observes in Fig.~\ref{fig:scaling}(a) that the thickness of the vapor film at the neck follows the scaling law $\tilde{h}_{f,\rm neck}\sim\mathcal{\tilde E}^{1/3}$ whatever the drop size. The same trend and exponent have been found for a Leidenfrost drop on a solid substrate~\cite{sobac2014}. The variation of the thickness of the vapor film at the neck with the drop radius shows different trends for small and large drops (see Fig.~\ref{fig:scaling}(c)). This observation is also similar to the solid-substrate case. Hence, the nature of the substrate does not seem to be crucial for the behavior of the thickness of the vapor film at the neck. However, this observation does not hold for its thickness at the center. On the one hand, for moderately small drop sizes $\tilde{R}\lesssim 1$, we notice that the tendencies shown in Fig. \ref{fig:scaling}(b,d) for the liquid substrate are still roughly the same as those found in \cite{sobac2014} for a solid substrate. In particular, we recover the typical scaling $\tilde{h}_{f,\rm center}\sim \tilde{\mathcal{E}}^{1/6}$ and even its drift to larger values of the exponent (eventually approaching $1/3$, the neck exponent) as $\tilde{R}$ is decreased and/or $\tilde{\mathcal{E}}$ is increased. For both substrate types, such a scaling variation goes together with the corresponding one in the vapor film shape: the vapor film uniformizes throughout, with its pronounced neck-pocket structure ceasing to exist. On the other hand, for larger sizes, $\tilde{R}\gtrsim1$, we have here predicted a drastic change of the geometry of the vapor pocket (see Fig.~\ref{fig:DropShapes}) between the cases of liquid and solid substrates. Quite correspondingly, for $\tilde{R}\gtrsim1$, the scalings of $\tilde{h}_{f,\rm center}$ in Fig.\ref{fig:scaling}(b,d) have nothing to do with those obtained in \cite{sobac2014}.

To provide some typical dimensional values, for a drop of ethanol of radius $R=1.2$ mm ($\tilde{R}=0.8$) over a pool of silicone oil V20 at $T_p=118^\circ$C ($\tilde{\mathcal{E}}=2.23\times10^{-7}$), the thicknesses of the vapor layer at the center and at the neck are respectively $h_{f,\rm center}=57 \, \rm{\mu m}$ and $h_{f,\rm neck}=18 \, \rm{\mu m}$.

\subsection{Evaporation dynamics}\label{sec:Mod_evap}

\begin{figure}[h!]
\centering
\begin{minipage}{0.45\textwidth}
\centering
(a)\\
		\includegraphics[height=5cm]{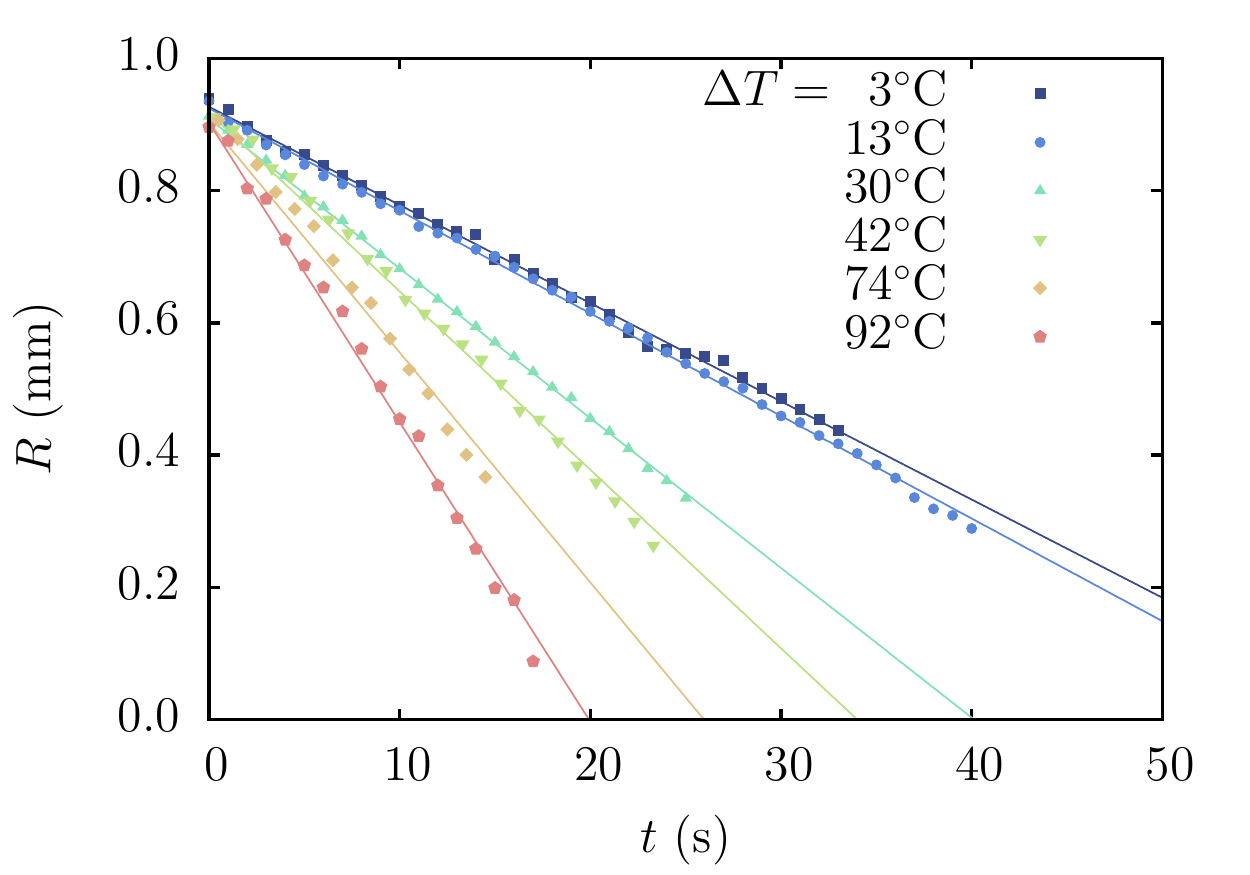}
\end{minipage}
\begin{minipage}{0.45\textwidth}
\centering
\vspace{0.5cm}
(b)\\
	\includegraphics[height=5cm]{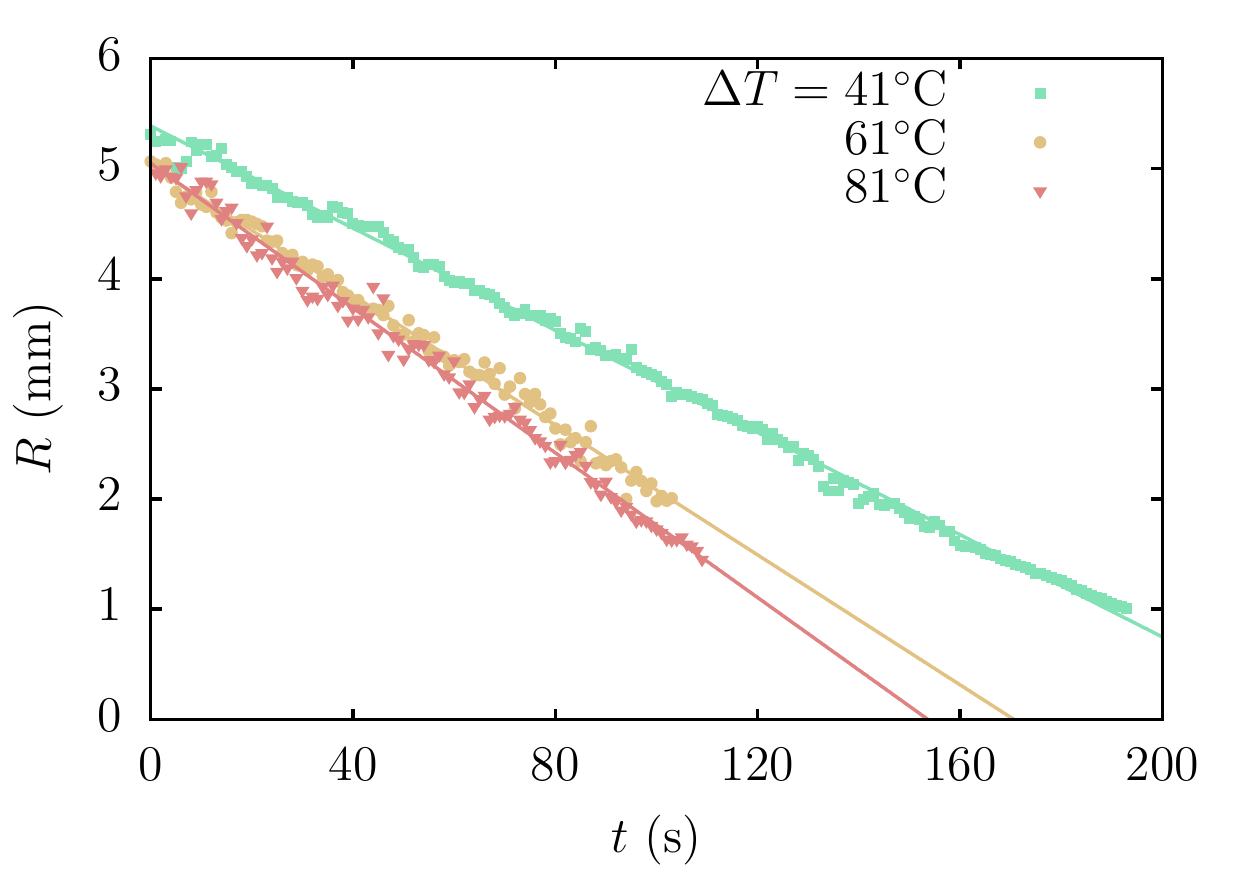}
\end{minipage}

 	\caption{Radius of an ethanol drop as a function of time on a pool of silicone oil V20 at different superheats (see legends). Plots (a) and (b) differ by the initial drop radius (small and large, respectively). The points are experimental data while the continuous lines are the fits by Eq.(\ref{eq:RVst_general}) with $\alpha=1$ (linear trend).}
		\label{figures:RVst_Ethanol}
\end{figure}

In order to characterize the drop evaporation process, we experimentally measure the time evolution of the drop radius over time (as viewed from above), which is illustrated in Fig.~\ref{figures:RVst_Ethanol}. Whereas for small drops data are provided for a wide range of superheats, for large drops data can only be provided for a narrower range of superheats since they are rather fragile at low superheats as explained in Sec.~\ref{sec:quali_obs}. Assuming a purely conductive heat transport across the vapor film and a lubrication flow in the film, previous studies~\cite{biance2003leidenfrost} have reported that the radius of Leidenfrost drop evaporating on a solid substrate follows the power law

\begin{equation}
R  (t) = R_0  \left( 1- \frac{t}{\tau} \right) ^\alpha \ ,
\label{eq:RVst_general}
\end{equation}

\begin{figure*}[ht!]
\centering
		\includegraphics[width=\textwidth]{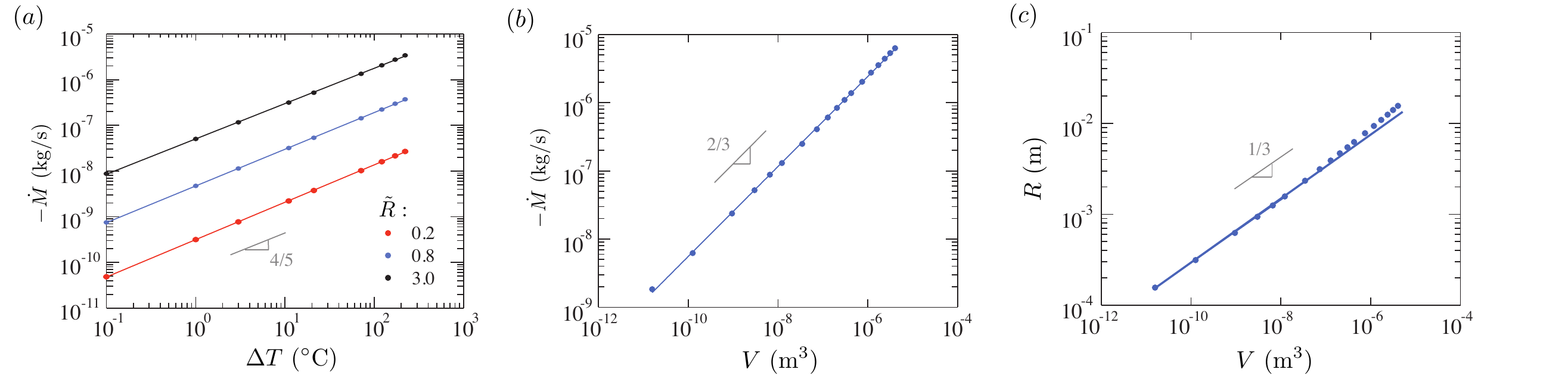}
 	\caption{Evaporation rate as a function of (a) the superheat for different drop sizes and (b) the drop volume for $\tilde{\mathcal{E}}=2.23\times10^{-7}$. (c) Link between the drop radius (as viewed from above) and the drop volume. Case considered: ethanol Leidenfrost drops over a silicone oil pool ($T_p=118^{\circ}$C).}
		\label{fig:Evaporation}
\end{figure*}

\noindent where $R_0$ is the initial radius of the drop, $\tau$ is the drop characteristic lifetime and $\alpha$ is an exponent depending on the shape of the drop. For a drop with a radius larger than the capillary length $\ell_{c_d}$ (puddle shape), $\alpha = 2$, and for drop with a radius smaller than $\ell_{c_d}$ (quasi-spherical shape), $\alpha = 0.5$. In our case of a Leidenfrost drop evaporating above a deformable substrate, the data still appear to be well fitted by Eq.(\ref{eq:RVst_general}). However, $\alpha = 1$ here whatever the value of the drop radius. Indeed, one sees in Fig.~\ref{figures:RVst_Ethanol} that the drop radius appears to decrease linearly over time.

Such a linear behavior is also predicted from our model. An integration of the local evaporation flux $\mathcal{J}$ over the vapor film (from $ r=0$ to $r= R_{\ast}$) yields $-\dot{M}=\frac{2 \pi \lambda_v \Delta T}{\mathcal{L}}\int_{0}^{R_{\ast}} \frac{r \ \text{d}r}{h-e}$, $M$ being the mass of the drop and the dot representing the time derivative, and where the equality to $(-\dot{M})$ implies neglecting contributions from the rest of the drop surface. As it is seen in this expression and shown in Fig.~\ref{fig:Evaporation}, the evaporation rate depends on the superheat and on the size of the drop. Figure~\ref{fig:Evaporation} reveals that $-\dot{M}\sim {\Delta T}^{4/5}$ whatever the drop size and $-\dot{M}\sim {V}^{2/3}$ where $V$ is the volume of the drop. Since $-\dot{M}= - \rho_{\ell} \ \text{d}V  /\text{d}t \sim {V}^{2/3} \, \Delta {T}^{4/5} $, we deduce that $V (t)\simeq V_0 (1-t /\tau)^{3}$ with $\tau \sim {V_0^{1/3}} / \Delta {T}^{4/5}$. From the relationship $R\sim V^{1/3}$ between the apparent drop radius and the drop volume (see Fig.~\ref{fig:Evaporation}(c)), one therefore approximately obtains that $R \simeq R_0 (1-t/\tau)^{\alpha}$ with $\alpha=1$. Indeed, this linear evolution of the drop radius during the drop evaporation appears to correctly match the experimental results presented in Fig.~\ref{figures:RVst_Ethanol}. One notices that the relation between the drop radius $R$ and its volume $V$ differs from the case of a solid substrate when the drop radius is larger than $\ell_{c_d}$. Indeed, in the case of a flat substrate, large drops adopt a puddle shape which verifies $V \propto {R} ^2$. Such a difference results from the fact that large drops over a liquid do not experience saturation in thickness, but continue to penetrate into the pool as observed in Fig.~\ref{fig:DropShapes}(a).

\begin{figure}[h!]
\centering
		\includegraphics[width=8cm]{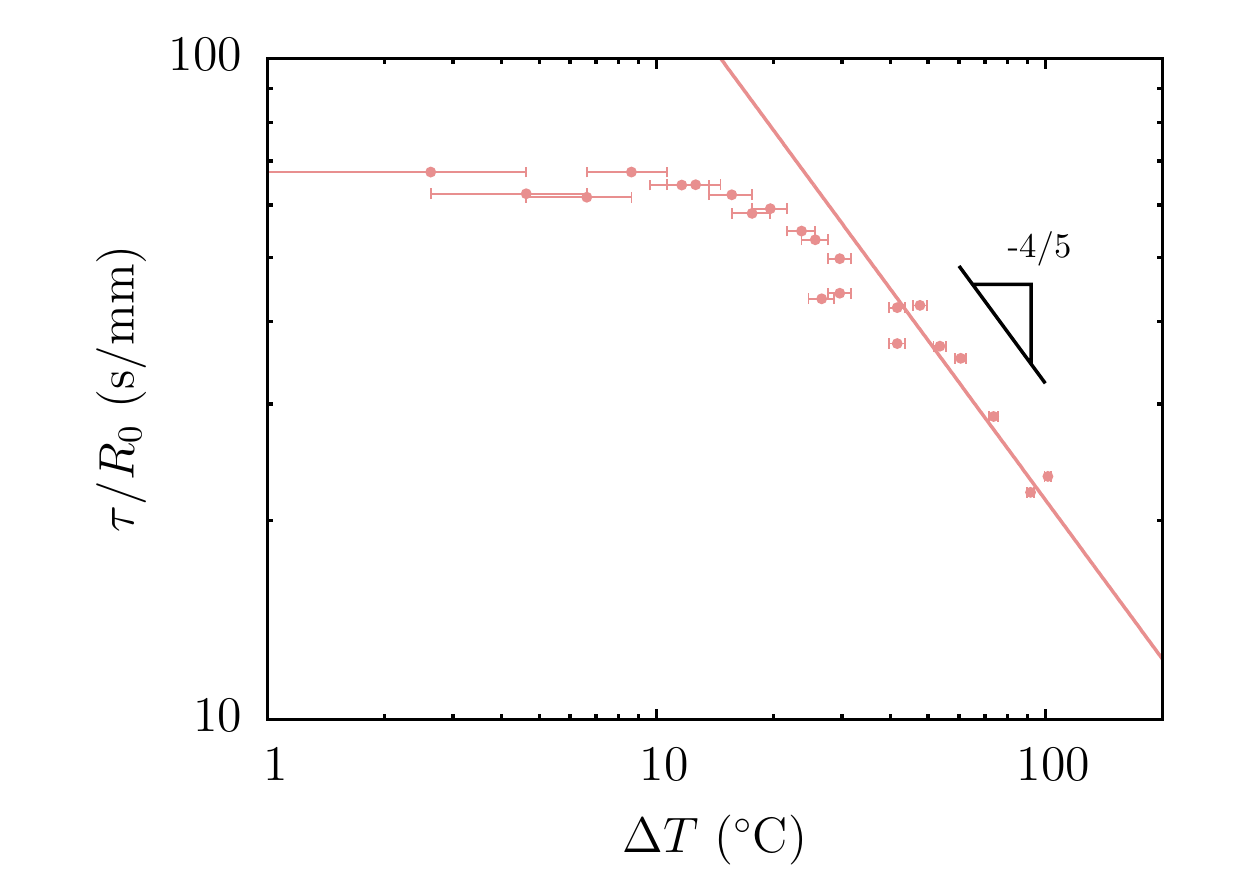}
 	\caption{Reduced drop lifetime $\tau/R_0$ as a function of the superheat $\Delta T$ for drops of ethanol and a pool of silicone oil V20. The solid line indicates the scaling law $\tau/R_0  \sim \Delta {T}^{-4/5} $.}
		\label{fig:droplifetime}
\end{figure}

The fits of the curves $R(t)$, such as shown in Fig.~\ref{figures:RVst_Ethanol}, by Eq.(\ref{eq:RVst_general}) allow to estimate the reduced lifetime of the drops $\tau/R_0$, which can serve as an indicator of the evaporation rate of the drops. The larger is $\tau/R_0$, the smaller is the evaporation rate. The variation of this quantity with the superheat $\Delta T$ is shown in Fig.~\ref{fig:droplifetime}. One observes that the larger is the superheat, the smaller is the reduced drop lifetime, as expected. The relative uncertainties estimated from the combination of the errors on the initial radius and on the drop characteristic lifetime are below $1$ percent. A continuous line shows the law $\tau/R_0 \propto \Delta T^{-4/5}$ which is explained above. However, the data clearly deviates from this law below $\Delta T = 30^\circ$C, and saturates below $\Delta T = 10^\circ$C.

The reason for the discrepancy at smaller superheats is not completely clear. However, it may be due to the fact that the evaporation from the upper part of the drop, neglected in the present analysis against the evaporation through the vapor film, becomes nonetheless relatively significant at smaller superheats. This evaporation, on the first order, depends on the gradient of temperature between the temperature of the drop $T_{\rm sat}$ and the room temperature, which does not depend on the temperature of the pool and thus explains the saturation at low superheats. This evaporation can be neglected when the superheat is large as the evaporation below the drop becomes significant, but not when the superheat is low. In the case of a solid substrate, this saturation cannot be observed because the Leidenfrost temperature is much higher than the crossover between the two regimes.

The death of a Leidenfrost drop on a hot liquid pool is different of the one on a solid substrate. Indeed, on a solid substrate, Celestini~\textit{et~al.} showed that a Leidenfrost drop experiences a take-off at the end of its lifetime~\cite{celestini2012take}. In our case, for a sufficiently small radius, the drop is rather observed to contact the pool, as already mentionned. Such a critical drop radius depends on the superheat and even on the initial radius. For instance, one observes in Fig.~\ref{figures:RVst_Ethanol}(a) that for $\Delta T = 3^\circ$C, the levitated state exists up to a radius of about $0.4\ $mm whereas for $\Delta T = 92^\circ$C, the drop survives in a levitated state up to a really small radius. However, the data is rather scattered and no quantitative tendency can be found (see supplementary material). One can notice that for a drop radius around $0.3\ $mm, and a superheat of $1^\circ$C ($\tilde{\mathcal{E}} \simeq 5 \times10^{-9}$), the thickness of the vapor layer at the neck becomes of the order of one micron (see Fig.~\ref{fig:scaling}) which becomes close to the case where van der Waals forces comes into play. Finally, the contact leads to a liquid lens that is very quickly evaporated.

\section{Further considerations}
\label{sec:considerations}

The presented model actually treats the substrate as an ideal deformable substrate, \textit{i.e.} the pool is assumed to be isothermal and no flow develops inside. As a consequence, the influence of the pool viscosity and thermal properties is not considered in the model. Therefore, even if our model appears to rather correctly capture the main mechanisms of the phenomenon, it has inherent limitations which will be discused in this section. The only properties of the pool accounted for in the model are its density and surface tension, whose influence will also be investigated.

\subsection{Effects of density and surface tension ratios}
\label{sec:Density/SurfaceTension}

\begin{figure}[h!]
\centering
		\includegraphics[width=8cm]{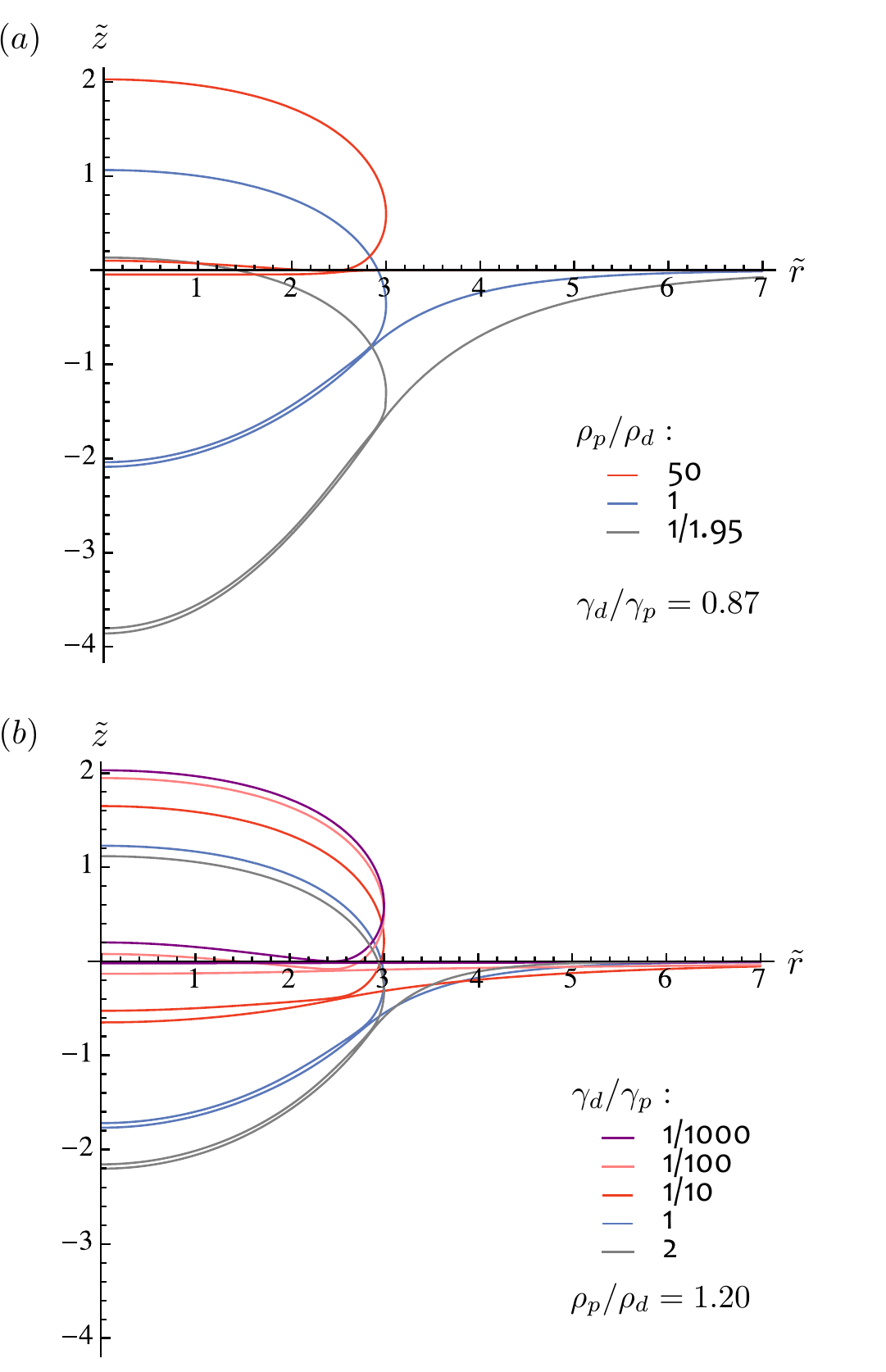}
 	\caption{Influence of the ratios of the liquid densities (a) and surface tensions (b) on Leidenfrost shapes for a drop of a size $\tilde R=3$ with $\mathcal{\tilde E}=2.23\times10^{-7}$ (corresponding to a drop of ethanol over a pool of silicone oil V20 at $T_p=118^{\circ}$C).}
		\label{fig:DensitySurfaceTension}
\end{figure}

Figures~\ref{fig:DensitySurfaceTension} present the results of a parametric study highlighting the role of the ratios of the densities and of the surface tensions of the liquids of the pool and of the drop on the geometry of the pool and drop surfaces. One notices (Fig.~\ref{fig:DensitySurfaceTension}) that an increase of ${{\rho_p}}/{{\rho_d}}$ or ${\gamma_p}/{\gamma_d}$ naturally makes the situation closer to a Leidenfrost drop on a flat surface. Of course, these changes also alter the evaporation, and, for very high values of ${{\rho_p}}/{{\rho_d}}$ or ${\gamma_p}/{\gamma_d}$, the evaporation flux also gets closer to that of a Leidenfrost drop on a flat solid substrate. However, practically, fluids densities and surface tensions are never sufficiently different to reach this extreme case. On the contrary, a decrease of these ratios leads to an increase of the pool deformation and thus a deeper immersion of the drop inside the pool.

Interestingly, large drops deeply immersed with a virtually flat surface at their top are also encountered in the case of delayed coalesce of large floating oil drop over a pool of the same liquid~\cite{couder2005bouncing}. In our study, for the case of the drops that are deeply immersed, we do not observe drops with their upper part fully below $z=0$, either experimentally or numerically. As an example of drops with a deeper immersion, we released drops of HFE--7100 over the same silicone oil pool as previously ($\frac{\gamma_d}{\gamma_p} = 0.7$ and $\frac{\rho_p}{\rho_d} =0.65$). The largest drops we were able to observe experimentally had a radius $\tilde R \simeq 2$, and they were always emerging above the pool base level.

\subsection{Influence of the pool viscosity}\label{sec:viscosity}

\begin{figure}[h!]
\centering
\begin{minipage}{0.45\textwidth}
\centering
\hspace{0.4cm}(a)\\
\vspace{0.cm}
		\includegraphics[width=8cm]{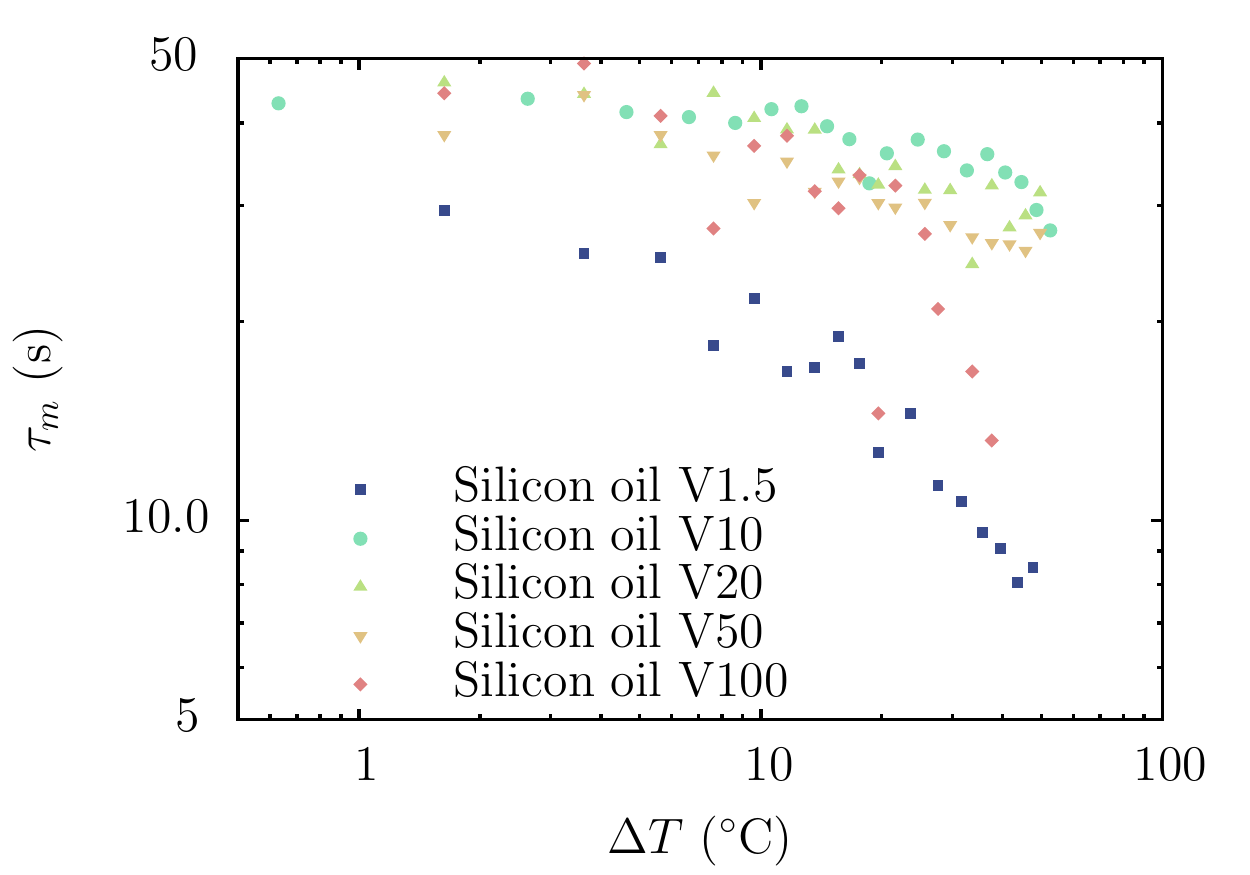}
\end{minipage}
\begin{minipage}{0.45\textwidth}
\centering
\vspace{0.5cm}
\hspace{1cm}(b)\\
	\includegraphics[width=8cm]{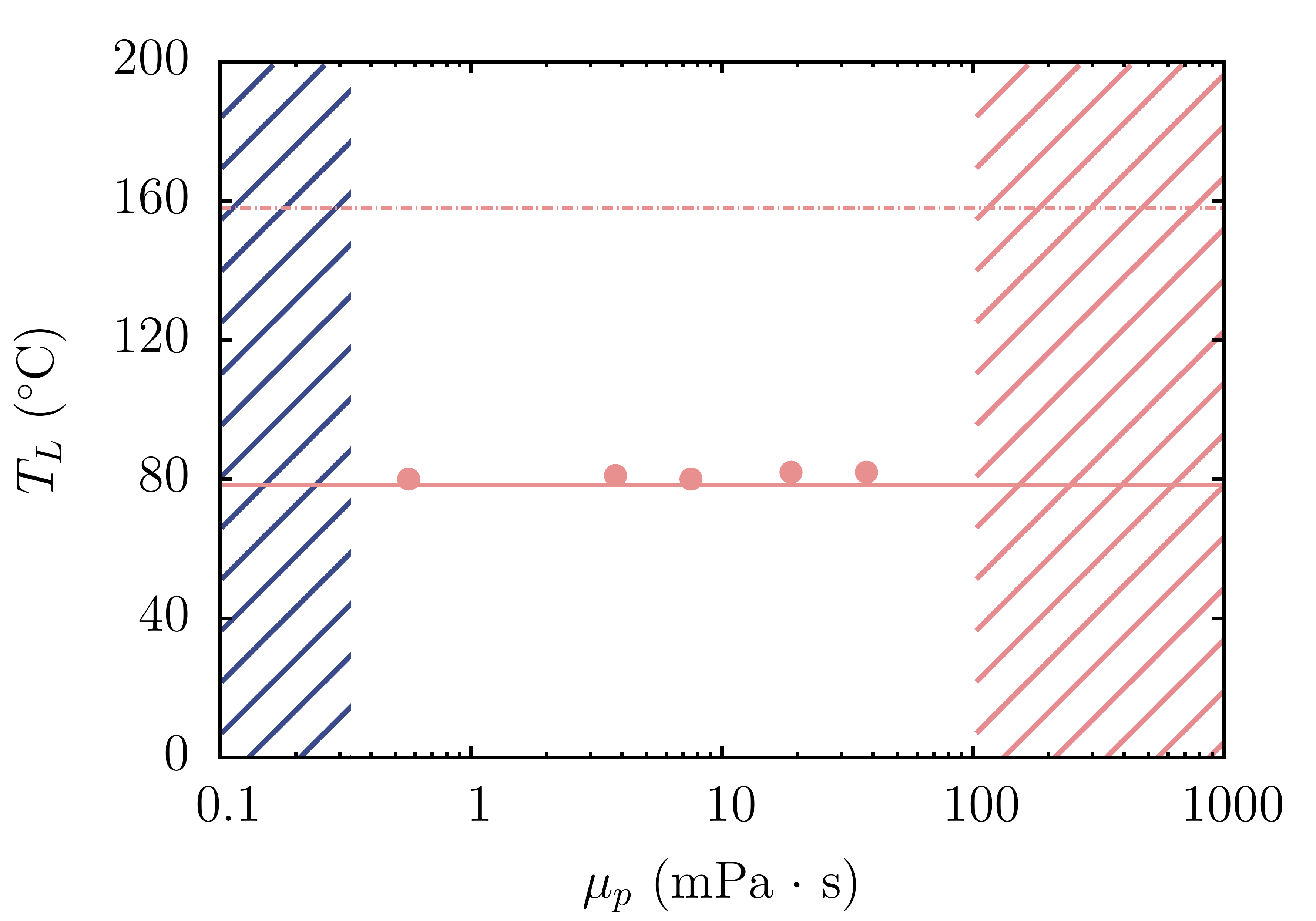}
\end{minipage}
 	\caption{(a) Maximum lifetime of ethanol Leidenfrost drops ($R_0 = 0.9$ mm) on several silicone oil pools as a function of the temperature of these substrates. (b) Associated Leidenfrost temperature $T_L $ of these drops as a function of the dynamic viscosity $\mu_p$ of the oil of the pool at the Leidenfrost temperature. The continuous and dashed lines represent the boiling temperature of the ethanol and its Leidenfrost point on an aluminum substrate respectively~\cite{wang2000critical}. The hatched blue zone is the zone where it is hard to find any liquid and the hatched pink zone is the zone where we did not observe stable Leidenfrost drops.}
		\label{figures/viscosity}
\end{figure}

We performed experiments similar to the case presented above with drops of ethanol on hot pools of silicone oils of various viscosities. A Leidenfrost effect has been observed for a dynamic viscosity of the pool $\mu_p$ up to $50$ $\rm{mPa \cdot s}$. In contrast, we could never observe a Leidenfrost effect for drops on a hot pool of silicone oil with a dynamic viscosity $\mu_p$ larger than $150\ \rm{mPa \cdot s}$ unless the liquid of the drop is heated at its boiling temperature before being deposited on the surface. This absence of Leidenfrost effect for highly viscous liquid substrates and the fact that the preheating helps the drops to stay in the Leidenfrost state strongly suggest that the convection in the pool is somehow important for the Leidenfrost state to be maintained. Supposedly, the convection in the pool may not be sufficient to prevent the cooling of its surface under the drop for highly viscous oils. In Fig.~\ref{figures/viscosity}(a), we plot the maximum lifetime $\tau_m$ observed for ethanol drops (initial radius $R_0 \simeq 0.9$ mm), \textit{i.e.} the maximum of the lifetimes of three different drops in the same conditions, as a function of the superheat $\Delta T$ for differents types of oil. This maximum lifetime decreases monotonically as the superheat increases. We also see that the curves are quite similar except for the substrate made of silicone oil~V1.5.

In Fig.~\ref{figures/viscosity}(b), we plot the associated Leidenfrost temperature $T_L $ for each of the pool liquids used as a function of their dynamic viscosity $\mu_p$. We also indicate the boiling temperature (solid line) as well as the Leidenfrost temperature for an ethanol drop on an aluminum substrate (dashed line), \textit{i.e.} $158^\circ$C~\cite{wang2000critical}. One observes that the Leidenfrost temperature is very low and practically does not vary in the range of the pool viscosities used. The hatched blue zone is the zone where very few liquids exist (\textit{e.g.} liquid nitrogen, $\mu = 0.018\ \rm{mPa\cdot s}$). In addition, most of these liquids have a low boiling point and cannot be used as a heated substrate for a Leidenfrost drop. The hatched zone in pink is the zone where we hardly observe any Leidenfrost drop. Therefore, we see that the Leidenfrost temperature is very stable for ethanol in large range of substrates viscosities. 

\subsection{Marangoni effect}
\label{sec:Marangoni}

Depositing a Leidenfrost drop over the surface of the pool leads to a local alteration of the heat transfer, which may result in a Marangoni flow in the pool. A similar phenomenon has been extensively studied by Savino~\textit{et~al.} in the case of a drop deposited on a slightly hotter pool (about $10^\circ$C)~\cite{savino2003marangoni}. They proved that the Marangoni flow can delay the drop coalescence for a few seconds.


However, in the case studied by Savino~\textit{et~al.}, the levitation never persists when the superheat reaches less than $3^\circ$C. Furthermore, the coalescence can be delayed in the best case for $30$ s, while our large drops can levitate up to $200$ s. 

Moreover, experiments that are not shown here show that there exists flows in the pool under the drop. However, depending on the liquid (ethanol or HFE--7100), the flow at the surface of the pool is directed either towards the drop or away from it (resp.). Thus, there exist cases in which the Marangoni effect in the pool is not sufficient to increase the pressure under the drop (\textit{e.g.} HFE--7100) and cannot be responsible for the levitation. However, as there exists some flows in the pool, further study is needed to fully clarify their impact on the phenomenon.

Considering these facts and the satisfying agreement between our model and the experimental data, the full consideration of the impacts of the Marangoni effect is beyond the scope of this paper, but should be consider in further studies.

\section{Conclusion}
\label{sec:conclusion}

We have experimentally and theoretically studied Leidenfrost drops placed over a hot liquid pool and stressed the importance of the nature of the substrate on the Leidenfrost effect. We showed experimentally that over a pool, a Leidenfrost state is possible as soon as the liquid of the pool is just hotter than the drop boiling point, with no apparent Leidenfrost threshold. This is at least partly due to the fact that a liquid substrate has no roughness unlike the solid ones. The early apparition of a Leidenfrost effect on a liquid pool is a phenomenon which is of primary importance regarding the extinction of pool fires and the spray cooling of hot baths. Experiments have also revealed that the final stage of the drop life is quite different over the liquid pool and solid substrates: instead of a take-off over the latter, a sufficiently small drop simply contacts onto the former. The maximum depth of the pool surface deformation predicted by the lubrication-type model is in agreement with measurements. Scaling laws concerning the shape of the vapor film are also predicted. Another difference between liquid and solid substrates, established theoretically, is that over the former no chimney formation is predicted for relatively large Leidenfrost drops of the size for which it is definitely expected over the latter, \textit{i.e.} above 4 times the drop capillary-length. The dynamics of evaporation of these drops has been studied and the developed model yields a reasonable agreement with experiments except for small superheats for which we show that the evaporation through the vapor film ceases to be the predominant one. The nature of the substrate also changes the shape of large drops, saturating in height over a solid substrate (puddles) but not over a liquid one, which is due to deformation of the pool surface itself under the weight of the drop. The effect of the ratios of the surface tensions and of the densities of the liquids involved is also tackled and these ratios are seen to lead to a more solid-like case at some extremes or to a more liquid-like case. This study leaves several questions open such as the nature of the death of Leidenfrost droplets over a liquid substrate and the influence of the viscosity of the pool and of the Marangoni flows therein. Finally, we hope that this work also opens the way to further studies such as the impact of volatile drops on a hot pool or the particular mobility of these drops and ways to control or exploit it.

\begin{acknowledgments}
This research has been funded by the Interuniversity Attraction Pole Programme (IAP 7/38 MicroMAST) initiated by the Belgian Science Policy Office. MB, SD and PC gratefully acknowledge financial support of Fonds de la Recherche Scientifique--FNRS (the first for his FRIA grant, the last two for their Senior Research Associate Positions), and ESA-BELSPO (PRODEX projects). AD thanks University of Li\`ege through the ARC Supercool grant for the financial support of this work. This research was also carried out under COST Action MP1106's umbrella.
\end{acknowledgments}

\end{document}